\documentclass[conference]{IEEEtran} 

\usepackage {graphicx}  
\usepackage {amssymb}
\usepackage {amsmath}
\usepackage {mathrsfs}
\usepackage{amsmath}   

\begin{document}
%
\title{Adaptive Linear Programming Decoding}
\author{\authorblockN{Mohammad H. Taghavi N. and Paul H. Siegel}\\
\authorblockA{ECE Department, University of California, San Diego\\
Email: (mtaghavi, psiegel)@ucsd.edu}
}
%
%

\maketitle

\begin{abstract}
Detectability of failures of linear programming (LP) decoding and its potential for improvement by adding new constraints motivate the use of an adaptive approach in selecting the constraints for the LP problem. In this paper, we make a first step in studying this method, and show that it can significantly reduce the complexity of the problem, which was originally exponential in the maximum check-node degree. We further show that adaptively adding new constraints, e.g. by combining parity checks, can provide large gains in the performance. 
\end{abstract}


%
\IEEEpeerreviewmaketitle

\newtheorem{definition}{Definition}
\newtheorem{theorem}{Theorem}
\newtheorem{lemma}{Lemma}
\newtheorem{claim}{Claim}
\newtheorem{algorithm}{Algorithm}
\newtheorem{corollary}{Corollary}
\newtheorem{conjecture}{Conjecture}
\newtheorem{remark}{Remark}
\section{Introduction}
Linear programming (LP) decoding, as an approximation to maximum-likelihood (ML) decoding, was proposed by Feldman \emph{et al.} \cite{Feldman thesis}. Many observations suggest similarities between the performance of LP and iterative message-passing methods, e.g. in \cite{LP and MSA}. For example, we know that the existence of low-weight pseudo-codewords degrades the performance of both methods (\cite{Vontobel}, \cite{Feldman thesis}). Therefore, it is reasonable to make use of the simpler geometrical structure of LP decoding to make predictions on the performance of message-passing algorithms.

On the other hand, there are differences which prevent us from making an explicit connection between these two approaches. For instance, given an LDPC code, adding additional parity checks that are satisfied by all the codewords can only improve LP decoding, while with message-passing algorithms, these parity checks may have a negative effect by introducing short cycles in the Tanner graph. This property of LP decoding allows improvements by tightening the relaxation. Another characteristic of LP decoding (the \emph{ML certificate property}) is that its failure to find the ML codeword is detectable. More specifically, the decoder always gives either the ML codeword, or a nonintegral pseudo-codeword as the solution. 

These two properties motivate the use of an adaptive approach in LP decoding which can be summarized as follows: Given a set of constraints that describe a code, start the LP decoding with a few of them, and sequentially and adaptively add more of the constraints to the problem until either the ML codeword is found or no further ``useful'' constraint exists. The goal of this paper is to explore the potential of this idea for LP decoding. 
We also prove some properties for the LP relaxation of ML decoding which can be useful for performance analysis of LP and/or iterative decoding algorithms.
We show that by putting LP in an adaptive setting, we can obtain the same performance as if a huge number of constraints were added to the relaxation from the beginning. In particular, we have observed that while the number of constraints per check node required for convergence is exponential in the check node degrees for LP decoding, the adaptive method generally converges with a number of constraints which is a (small) constant, independent of degree distributions. This property makes it feasible to apply LP decoding to higher-density codes. 

The rest of this paper is organized as follows. In Section II, we review Feldman's LP decoding. In Section III, we introduce and analyze an adaptive algorithm to solve the original LP problem more efficiently. In Section IV, we study how adaptively adding additional constraints can improve the performance. Finally, Section V concludes the paper. Due to space limitations, some of the proofs of the results are omitted.
%

\section{LP Relaxation of ML Decoding}
Consider a binary linear code $\mathscr{C}$ of length $n$. If a codeword $\underline y\in \mathscr{C}$ is transmitted through a binary-input memoryless channel, the ML codeword given the received vector $\underline r \in \mathbb{R}^n$ is the solution to the optimization problem
\begin{eqnarray}
\label{ML decoding}
\textrm{minimize} &&\underline \gamma^T \underline x \nonumber\\
\textrm{subject to} &&\underline x\in \mathscr{C},
\end{eqnarray}
where $\underline \gamma$ is the vector of log-likelihood ratios defined as
\begin{equation}
\label{def gamma}
\gamma_i= \log \left({\Pr(r_i|y_i=0) \over \Pr(r_i|y_i=1)} \right).
\end{equation}
As an approximation to ML decoding, Feldman \emph{et al.} proposed a relaxed version of this problem by first considering the convex hull of the local codewords defined by each row of the parity-check matrix, and then intersecting them to obtain what is called the \emph{fundamental polytope} by Koetter \emph{et al.} \cite{Vontobel}. This polytope has a number of integral and nonintegral vertices, but the integral vertices exactly correspond to the codewords of $\mathscr{C}$. Therefore, whenever LP gives an integral solution, it is guaranteed to be the ML codeword. 

In Feldman's relaxation of the decoding problem, the following is done for each row $j=1,\ldots,m$ of the parity-check matrix.
Suppose that the $j$th check node has the neighborhood set $N\subset \{1, 2, \ldots, n\}$, i.e. $N$ contains the indices of the variable nodes that are directly connected to this check node. Then, add the following constraints to the problem:
\begin{equation}
\label{constraints}
\sum_{i\in V} x_i -\!\!\!\sum_{i\in N\backslash V}\!\!\! x_i \leq |V|-1,\ \forall\ V\subset N\ \textrm{such that}\ |V|\ \textrm{is odd}.
\end{equation}
Throughout the paper, we refer to the constraints of this form as \emph{parity-check constraints}. In addition, for any element $x_i$ of the optimization variable, $\underline x$, the constraint that $0\leq x_i \leq 1$ is also added.

\section{Adaptive LP Decoding}
As any odd-sized subset $V$ of the neighborhood $N$ of each check node introduces a unique parity-check constraint, there are $2^{d_c-1}$ constraints corresponding to each check node of degree $d_c$. Therefore, the total number of constraints and hence, the complexity of the problem, is exponential in terms of the maximum check node degree, $d_c^{max}$. This becomes more significant in a high density code where $d_c^{max}$ increases with the code length, $n$. In this section, we show that LP relaxation of linear codes has some properties that allow us to solve the optimization by using a much smaller number of constraints.

\subsection{Properties of the Relaxation Constraints}
\begin{definition}
Given a constraint of the form
\begin{equation}
\label{general const}
\underline a_i^T \underline x\leq b_i,
\end{equation}
and a vector $\underline x_0 \in \mathbb{R}^n$, we call (\ref{general const}) an \emph{active constraint} at $\underline x_0$ if
\begin{equation}
\label{active const}
\underline a_i^T \underline x_0=b_i,
\end{equation}
and a \emph{violated constraint} or, equivalently, a \emph{cut} at $\underline x_0$ if
\begin{equation}
\label{violated const}
\underline a_i^T \underline x_0>b_i.
\end{equation}
\end{definition}

Considering a constraint that generates a cut
\begin{equation}
\label{typical cut}
\sum_{i\in V} x_i -\sum_{i\in N\backslash V} x_i > |V|-1,
\end{equation}
at point $\underline x$, we can immediately make the following observations:
\begin{equation}
\label{obsv1}
|V|-1 < \sum_{i\in V} x_i \leq |V|,
\end{equation}
\begin{equation}
\label{obsv2}
0\leq \sum_{i\in N\backslash V}\!\!\! x_i < x_j\ \ \forall j\in V.
\end{equation}
The following theorem reveals a special property of the constraints of the LP decoding problem.
\begin{theorem}
\label{one cut per check}
At any given point $\underline x \in [0,1]^n$, at most one of the constraints introduced by each check node can be a cut. (Proof omitted.)
\end{theorem}

Having a linear $(n, k)$ code with $m=n-k$ parity checks, a natural question is how we can find all the cuts defined by the LP relaxation at any given point $x\in \mathbb{R}^n$. For any check node and an odd subset $V$ of its neighborhood that introduces a cut, we know from (\ref{obsv2}) that the members of $V$ are the variable nodes with the largest values among the neighbors of the check node. Therefore, sorting the elements of $\underline x$ before searching for a cut can simplify the procedure. 

Consider a check node $j$. Without loss of generality, assume that variable nodes $1, 2, \ldots , |N|$, are the neighbors of this check node, and they are sorted with respect to their values such that $x_1\geq x_2 \geq \cdots \geq x_{|N|}$. The following algorithm is an efficient way to find the cut generated by this check node at $\underline{x}$, if it exists.

\begin{algorithm}
\label{find cuts}
\end{algorithm}

\emph{Step 1: Set $v=1$, $V=\{1\}$ and $V^c\triangleq N\backslash V=\{2, 3, \ldots, |N|\}$.}

\emph{Step 2: Check the constraint (\ref{constraints}). If it is violated, we have found the cut. Exit.}

\emph{Step 3: Set $v=v+2$. If $v\leq |N|$, move $x_{v-1}$ and $x_{v}$ (the two largest members of $V^c$) from $V^c$ to $V$}

\emph{Step 4: If $v\leq |N|$ and (\ref{obsv1}) is satisfied, go to Step 2; otherwise, the check node does not provide a cut at $x$.}

If redundant calculations are avoided, this algorithm can find the cut generated by the check node, if it exists, in $O(d_c)$ time, where $d_c=|N|$ is the degree of the check node. Repeating the procedure for each check node, and considering $O(n\log n)$ complexity for sorting $\underline x$, the time required to find all the cuts at point $\underline x$ becomes $O(m d_c^{max} + n\log n)$ 
\footnote[1]{For low density codes, it is better to sort the neighbors of each check node separately, so the total complexity becomes $O(m d_c^{max} + m d_c^{max}\log d_c^{max})$.}
.

\subsection{The Adaptive Procedure}
The Simplex LP algorithm starts from a vertex of the problem polytope and visits different vertices of the polytope by traveling through the edges until it finds the optimum vertex. The time required to find the solution is approximately proportional to the number of vertices that have been visited, and this, in turn, is determined by the number and properties of the constraints in the problem. Hence, if we eliminate some of the intermediate vertices and only keep those which are close to the optimum point, we can reduce the complexity of the algorithm. To implement this idea in the adaptive LP decoding scheme, we run the LP solver with a minimal number of constraints to ensure boundedness of the solution, and depending on the LP solution, we add only ``the useful constraints'' that cut the current solution from the feasible region. This procedure is repeated until no further cut exists.

To start the procedure, we need at least $n$ constraints so that the problem has a vertex that can become the solution of the first round. Using the condition that $0\leq x_i \leq 1$, we add one side of these inequalities for each $i$, depending on whether increasing $x_i$ increases or decreases the objective function. In other words, for each $i\in \{1, 2, \ldots, n\}$, we initially have the constraint
\begin{eqnarray}
\label{initial constraints}
0\leq x_i \ \ \textrm{if}\ \gamma_i>0, \nonumber \\
x_i\leq 1 \ \ \textrm{if}\ \gamma_i<0.
\end{eqnarray}
The optimum (and only) vertex of this initial problem corresponds to the result of (uncoded) hard-decision based on the received vector. Now we proceed with the following algorithm:

\begin{algorithm}
\label{adaptive LP}
\end{algorithm}

\emph{Step 1: Setup the initial problem according to (\ref{initial constraints}).}

\emph{Step 2: Run the LP solver.}

\emph{Step 3: Search for all the cuts for the current solution.}

\emph{Step 4: If one or more cuts are found, add them to the problem constraints and go to Step 2.}

\begin{claim}
If at any iteration of Algorithm \ref{adaptive LP} no cut is found, the current solution is the the solution of the LP decoder with all the relaxation constraints given in Section II.
\end{claim}
\begin{proof}
The claim follows from the fact that if at any stage no cut is found, the current solution is in the fundamental polytope. 
\end{proof}

\subsection{A Bound on the Complexity}
\begin{theorem}
\label{n iterations}
The adaptive algorithm (Algorithm \ref{adaptive LP}) converges with at most $n$ iterations.
\end{theorem}
\begin{proof}
The final solution is a vertex $\underline x_f$ of the problem space determined by the initial constraints along with those added by the adaptive algorithm. Therefore, we can find $n$ such constraints, $\kappa_i:\ \underline \alpha_i^T \underline x \leq \beta_i,\ i=1, 2, \ldots n,$ whose corresponding hyperplanes uniquely determine this vertex. This means that if we set up an LP problem with only those $n$ constraints, the optimal point will be $\underline x_f$. Now, consider the $k$th intermediate solution, $\underline x_k$, that is cut off at the end of the $k$th iteration. At least one of the constraints, $\kappa_1, \ldots, \kappa_n$, should be violated by $\underline x_k$, otherwise since $\underline x_k$ has a lower cost than $\underline x_f$, $\underline x_k$ would be the solution of LP with these $n$ constraints. But we know that the cuts added at the $k$th iteration are all the possible constraints that are violated at $\underline x_k$. Consequently, at least one of the cuts added at each iteration should be among $\{\kappa_i\}$
; hence, the number of iterations is at most $n$.
\end{proof}
\begin{remark}
This theorem applies to any general LP problem where there is a fixed set of constraints, and at each iteration we add all the cuts generated by this set of constraints.
\end{remark}
\begin{corollary}
The adaptive algorithm has at most $n(m+1)$ constraints at the final iteration.
\end{corollary}
\begin{proof}
Follows from Theorem \ref{one cut per check} and Theorem \ref{n iterations}.
\end{proof}

For high-density codes of fixed rate, this bound guarantees convergence with $O(n^2)$ constraints, whereas the standard LP and the polytope given in \cite{Feldman thesis} for high-density codes respectively require exponential and $O(n^3)$ constraints. 

\subsection{Numerical Results}
To observe the complexity reduction due to the adaptive approach for LP decoding, we have performed simulations over random regular LDPC codes of various lengths, degrees, and rates on the AWGN channel. All the experiments were performed with the low SNR value of $-1.0$ dB, since in the high SNR regime the recieved vector is likely to be close to a codeword, in which case the algorithm converges fast, rather than demonstrating its worst-case behavior. 

\begin{figure}
\centering
\includegraphics[width=3.3 in] {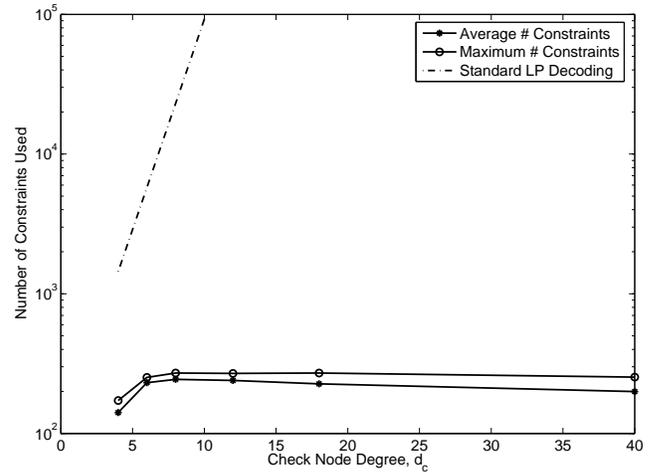}
\caption{The average and maximum number of parity-check constraints used versus check node degree, $d_c$, for fixed length $n=360$ and rate $R={1\over 2}$.}\!\!\!\!\!\!\!\!
\label{const vs d_c}
\end{figure}

In the first scenario, we studied the effect of changing the check node degree $d_c$ from $4$ to $40$ while keeping the code length at $n=360$ and the rate at $R={1\over 2}$. The simulation was performed over 400 blocks for each value of $d_c$. The maximum (average) number of iterations required to converge started from around $30$ ($14.5$) for $d_c=4$, and decreased monotonically down to $9$ ($5.9$) for $d_c=40$. The average and maximum numbers of parity-check constraints in the final iteration of the algorithm are plotted in Fig. \ref{const vs d_c}. It is observed that both the average and the maximum values are almost constant, and remain below $270$ for all the values of $d_c$. For comparison, the total number of constraints required for the standard (non-adaptive) LP decoding problem, which is equal to $2^{d_c-1}$ is also included in this figure. The decrease in the number of required constraints translates to a large gain for the adaptive algorithm in terms of the running time. This gain increases exponentially with $d_c$, so that with the LP solver that we have used in our work (GLPK \cite{GLPK}), the adaptive algorithm always converges several thousand times faster that standard LP for $d_c=8$.

\begin{figure}
\centering
\includegraphics[width=3.3 in] {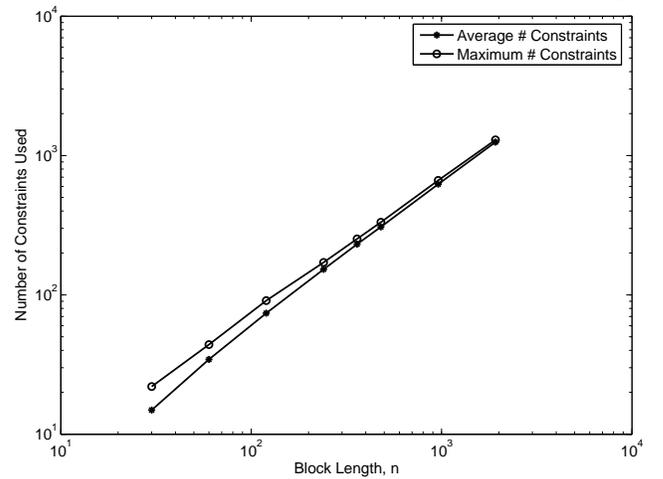}
\caption{The average and maximum number of parity-check constraints used versus block length, $n$, for fixed rate $R={1\over 2}$ and check node degree $d_c=6$.}\!\!\!\!\!\!\!\!
\label{const vs n}
\end{figure}

In the second case, we studied random (3,6) codes of lengths $n=30$ to $n=1920$. For all values of $n$, the maximum (average) number of required iterations remained between $10$ and $16$ ($5$ and $11$). The average and maximum numbers of parity-check constraints in the final iteration are plotted versus $n$ in Fig. \ref{const vs n}. We observe that the number of constraints is generally between $0.6 n$ and $0.7 n$. 

\begin{figure}
\centering
\includegraphics[width=3.3 in] {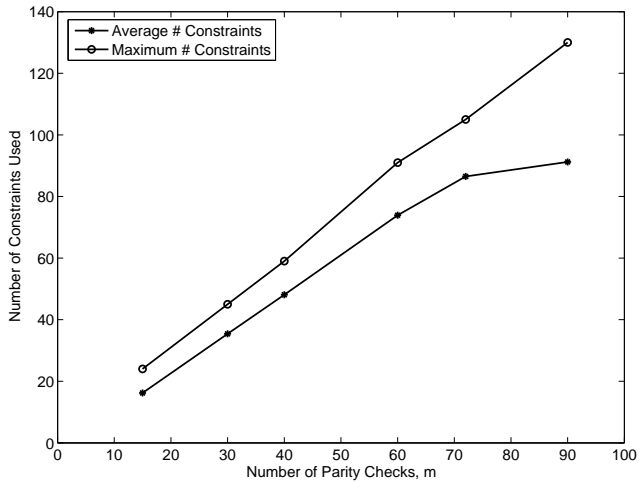}
\caption{The average and maximum number of parity-check constraints used versus the number of parity checks, $m$, for $n=120$ and $d_v=3$.}\!\!\!\!\!\!\!\!
\label{const vs m}
\end{figure}

In the third experiment, we investigated the effect of the rate of the code on the performance of the algorithm. Fig. \ref{const vs m} shows the average and maximum numbers of parity-check constraints in the final iteration where the block length is $n=120$ and the number of parity checks, $m$, increases from $15$ to $90$. The variable node degree is fixed at $d_v=3$. We see that the average and maximum numbers of constraints are respectively in the ranges $1.1m$ to $1.2m$ and $1.4m$ to $1.6m$ for most values of $m$. The relatively large drop of the average number for $m=90$ with respect to the linear curve can be explained by the fact that at this value of $m$ the rate of failure of LP decoding was less than $0.5$ at $-1.0$ dB, whereas for all the other values of $m$, this rate was close to $1$. Since the success of LP decoding generally indicates proximity of the received vector to a codeword, we expect the number of parity checks required to converge to be small in such a case, which decreases the average number of constraints.

Based on these simulation results, we observe that in practice the algorithm performs much faster than is guaranteed by Theorem \ref{n iterations}. These observations suggest the following conjecture. 
\begin{conjecture}
\label{conjecture on complexity}
For a random parity-check code of length $n$ with $m=n(1-R)$ parity checks and arbitrary degree distributions, as $n$ increases, the adaptive LP decoding algorithm converges with probability arbitrarily close to $1$ in at most $\alpha$ iterations and with at most $\beta$ final parity-check constraints per check node, where $\alpha$ and $\beta$ are constants independent of the length, rate and degree distribution of the code.
\end{conjecture}

\begin{remark}
\label{int positions in PCW}
If for a given code of length $n$, the adaptive algorithm converges with at most $q<n$ final parity-check constraints, then each pseudo-codeword of this LP relaxation should have at least $n-q$ integer elements. To see this, note that each pseudo-codeword corresponds to the intersection of at least $n$ active constraints. If the problem has at most $q$ parity-check constraints, then at least $n-q$ constraints of the form $x_i\geq 0$ or $x_i\leq 1$ should be active at each pseudo-codeword, which means that at least $n-q$ positions of the pseudo-codeword are integer-valued.
\end{remark}

\section{Generating Cuts to Improve the Performance}
The complexity reduction obtained by adaptive LP decoding inspires the use of cutting-plane techniques to improve the error rate performance of the algorithm. Specifically, when LP with all the original constraints gives a nonintegral solution, we try to cut the current solution, while keeping all the possible integral solutions (codewords) feasible.  

In the decoding problem, the new cuts can be chosen from a pool of constraints describing a relaxation of the maximum-likelihood problem which is tighter than the fundamental polytope. In this sense, the cutting-plane technique is equivalent to the adaptive LP decoding of the previous section, with the difference that there are more constraints to choose from. The effectiveness of this method depends on how closely the new relaxation approximates the ML decoding problem, and how efficiently we can search for those constraints that introduce cuts. Feldman \emph{et al.} \cite{Feldman thesis} have mentioned some ways to tighten the relaxation of the ML decoding, including adding redundant parity checks (RPC), and using lift-and-project methods. (For more on lift-and-project, see \cite{lift-and-project} and references therein.) Gomory's algorithm \cite{Gomory} is also one of the most well-known techniques for general integer optimization problems, although it suffers from slow convergence. Each of these methods can be applied adaptively in the context of cutting-plane techniques.

The simple structure of RPCs makes them an interesting choice for generating cuts. There are examples where even the relaxation obtained by adding all the possible RPC constraints does not guarantee convergence to a codeword. In other words, it is possible that we obtain a nonintegral solution for which there is no RPC cut, although the general case is still not well-studied. Also, finding efficient methods to search for RPC cuts for a given nonintegral solution is an open issue. On the other hand, as observed in simulation results, RPC cuts are generally strong, and a reasonable number of them makes the resulting LP relaxation tight enough to converge to an integer-valued solution. In this work, we focus on cutting-plane algorithms that use RPC cuts.

\subsection{Finding Redundant Parity-Check Cuts}
A redundant parity check is obtained by modulo-2 addition of some of the rows of the parity-check matrix, and this new check introduces a number of constraints that may include a cut. There is an exponential number of RPCs that can be made this way, and in general, most of them do not introduce cuts. Hence, we need to find the cuts efficiently by exploiting the particular structure of the decoding problem. In particular, we observe that cycles in the graph have an important role in determining whether an RPC generates a cut. This property is explained by Theorem \ref{RPC cycles}. 

\begin{definition}
\label{minimal subset}
Given a current solution, $\underline x$, the subset $T \subset \{1, 2, \ldots, m\}$ of check node indices is called a \emph{cut-generating collection} if the RPC made by modulo-2 addition of the parity checks corresponding to $T$ introduces a cut. 
\end{definition}

\begin{theorem}
\label{RPC cycles}
Let $T\subset \{1, 2, \ldots, m\}$ be a collection of check node indices in the Tanner graph of the code, and let $G$ be the subgraph made up of these check nodes, the variable nodes directly connected to them, and all the edges that connect them. Then, $T$ can be a cut-generating collection only if $G$ contains a cycle that only passes through variable nodes whose corresponding current values are fractional. (Proof omitted.)
\end{theorem}

This result motivates the following algorithm to search for cuts.

\begin{algorithm}
\label{search for RPC cuts}
\end{algorithm}

\emph{Step 1: Having a solution $\underline x$, prune the Tanner graph by removing all the variable nodes with integer values.}

\emph{Step 2: Starting from an arbitrary check node, randomly walk through the pruned graph until a cycle is found.}

\emph{Step 3: Create an RPC by combining the rows of the parity-check matrix corresponding to the check nodes in the cycle.}

\emph{Step 4: If this RPC does not introduce a cut, go to Step 2.}

By exploiting some of the properties of the problem as described above, the search for each cut becomes much faster than the naive search; however, this procedure is still the complexity bottle-neck of the decoding, and it becomes prohibitively complex for codes of length on the order of several hundred bits and above. On the other hand, experiments demonstrate that the number of cuts required to converge to the ML codeword (if the convergence is possible) does not grow very rapidly with length. This motivates more study of the properties of RPC cuts and efficient schemes to find them.

\begin{figure}
\centering
\includegraphics[width=3.3 in] {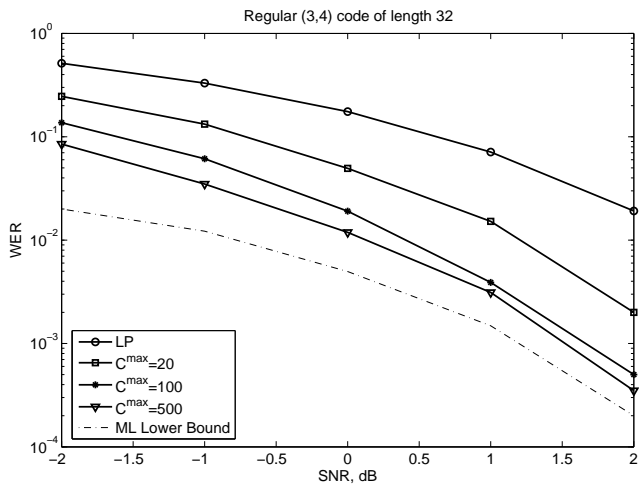}
\caption{WER of cutting-plane LP versus SNR for different values of $C_{max}$.}\!\!\!\!\!\!\!\!
\label{RPC WER}
\end{figure}

\subsection{Numerical Results}
To demonstrate the performance improvement achieved by using the RPC cutting-plane technique, we present simulation results for a random regular $(3,4)$ LDPC code of length $32$. The adaptive nature of the algorithm allows us to smoothly trade complexity for performance by changing the number of trials in the search for each RPC cut, i.e. the number of iterations in Algorithm \ref{search for RPC cuts}, as well as the total number of calls of the LP solver. In this experiment, we fix the total number of iterations of Algorithm \ref{search for RPC cuts} to $C_{max}$, and declare decoding failure if no cut is found after $C_{max}$ trials. 

The word error rate (WER) of the algorithm is plotted versus SNR in Fig. \ref{RPC WER} for different values of $C_{max}$. For comparison, the WER of pure LP decoding, i.e. with no RPC cut, and a lower bound on the WER of the ML decoder have been included, as well. In order to obtain this lower bound, we count the number of times that the cutting-plane LP algorithm with a large value of $C_{max}$ converges to a codeword other than the transmitted codeword, and divide that by the number of blocks. Due to the ML certificate property of LP decoding, we know that ML decoding would fail in those cases, as well. On the other hand, ML decoding may also fail in some of the cases where LP decoding does not converge to an integral solution. Therefore, this estimate gives a lower bound on the WER of ML decoding.

The numerical results suggest that the cutting-plane LP decoding with RPC cuts can significantly outperform the pure LP decoding, at the cost of increased complexity.

\section{Conclusion}
In this work, we studied the potential for improving LP decoding, both in complexity and error correction capability, by using an adaptive approach. The key idea was to use the fact that we can always recognize the failure of LP decoding to find the ML codeword, a property that message-passing algorithms only have in specific cases such as the erasure channel. This feature allows us to add only constraints that are ``useful'', depending on the current status of the algorithm.

In the algorithm proposed in Section III, the complexity is significantly reduced and becomes independent of the degree distributions, making it possible to apply LP decoding to parity-check codes of arbitrary densities. However, since general purpose LP solvers are used at each iteration, the complexity in terms of the block length is still super-linear, as opposed to linear as in the message-passing algorithms. An interesting question is whether we can design special LP solvers for decoding of LDPC codes that can take advantage of the sparsity of the constraints and other properties of the problem to converge in linear time.

Section IV serves as a first step to explore the application of cutting-plane techniques in LP decoding. A desirable feature of this approach is that by changing the parameters of the algorithm we can smoothly trade complexity for performance. In contrast, if we want to get the same performance gains by tightening the relaxation in a non-adaptive setting, the required complexity increases much faster. We showed that redundant parity checks provide strong cuts, even though they may not guarantee ML performance. A major open problem is to find efficient ways to search for these cuts by exploiting their properties, and to determine specific classes of codes for which RPC cuts are more effective. Furthermore, the effectiveness of cuts generated by other techniques, such as lift-and-project cuts and Gomory cuts, as well as specially designed cuts, needs further study.  
 

\appendices

\end{document}